\documentclass[twocolumn]{aastex62}

\graphicspath{{./}{figures/}}

\usepackage{amsmath, amsfonts, amssymb, bm} % math stuff
\usepackage{natbib} % use natbib
\received{November ??, 2018}
\revised{November ??, 2018}
\shorttitle{Magnetized fingering convection}
\shortauthors{Harrington \& Garaud}
\begin{document}
\turnoffedit

\title{Enhanced mixing in magnetized fingering convection, and implications for RGB stars}

\author{Peter Z. Harrington}
\affiliation{Department of Applied Mathematics, Baskin School of Engineering, University of California Santa Cruz, 1156 High Street, Santa Cruz, CA 95064, USA}

\author{Pascale Garaud}
\affiliation{Department of Applied Mathematics, Baskin School of Engineering, University of California Santa Cruz, 1156 High Street, Santa Cruz, CA 95064, USA}

%% Note that the \and command from previous versions of AASTeX is now
%% depreciated in this version as it is no longer necessary. AASTeX 
%% automatically takes care of all commas and "and"s between authors names.

%% AASTeX 6.2 has the new \collaboration and \nocollaboration commands to
%% provide the collaboration status of a group of authors. These commands 
%% can be used either before or after the list of corresponding authors. The
%% argument for \collaboration is the collaboration identifier. Authors are
%% encouraged to surround collaboration identifiers with ()s. The 
%% \nocollaboration command takes no argument and exists to indicate that
%% the nearby authors are not part of surrounding collaborations.

%% Mark off the abstract in the ``abstract'' environment. 
\begin{abstract}

Double-diffusive convection has been well studied in geophysical contexts, but detailed investigations of the regimes characteristic of stellar or planetary interiors have only recently become feasible. Since most astrophysical fluids are electrically conducting, it is possible that magnetic fields play a role in either enhancing or suppressing double-diffusive convection, but to date there have been no numerical investigations of such possibilities. Here we study the effects of a vertical background magnetic field (aligned with the gravitational axis) on the linear stability and nonlinear saturation of fingering (thermohaline) convection, through a combination of theoretical work and direct numerical simulations (DNSs). We find that a vertical magnetic field rigidifies the fingers along the vertical direction which has the remarkable effect of enhancing vertical mixing. We propose a simple analytical model for mixing by magnetized fingering convection, and argue that magnetic effects may help explain discrepancies between theoretical and observed mixing rates in low-mass red giant branch (RGB) stars. Other implications of our findings are also discussed.

\end{abstract}

%% Keywords should appear after the \end{abstract} command. 
%% See the online documentation for the full list of available subject
%% keywords and the rules for their use.

%% From the front matter, we move on to the body of the paper.
%% Sections are demarcated by \section and \subsection, respectively.
%% Observe the use of the LaTeX \label
%% command after the \subsection to give a symbolic KEY to the
%% subsection for cross-referencing in a \ref command.
%% You can use LaTeX's \ref and \label commands to keep track of
%% cross-references to sections, equations, tables, and figures.
%% That way, if you change the order of any elements, LaTeX will
%% automatically renumber them.
%%
%% We recommend that authors also use the natbib \citep
%% and \citet commands to identify citations.  The citations are
%% tied to the reference list via symbolic KEYs. The KEY corresponds
%% to the KEY in the \bibitem in the reference list below. 

\section{Introduction} \label{sec:intro}

Recent progress in quantifying transport by fingering (thermohaline) convection in stellar astrophysics \citep[see the review by][]{doi:10.1146/annurev-fluid-122316-045234} has reopened the debate on the origin of post dredge-up surface abundance variations in Red Giant Branch (RGB) stars. It has been known for a long time that stellar models which ignore non-canonical mixing do not predict any evolution in the surface abundances on the RGB after the first dredge up, but this prediction is at odds with observations \citep{2000A&A...354..169G}. More specifically, lithium and CNO cycle by-product abundances are observed to continue evolving with time, especially at the time of the so-called luminosity bump, which corresponds to the point at which the hydrogen-burning shell begins to expand into the region that was previously chemically mixed by the dredge-up event. 

\citet{2007A&A...467L..15C} proposed that fingering convection could be a natural explanation for this phenomenon, and would arise from the inverse $\mu$-gradient caused by the reaction 2He$^3 \rightarrow $He$^4 + 2p$ at the outer edge of the hydrogen-burning shell. Including the effect of fingering convection in their stellar evolution code using the mixing prescriptions of \citet{1972ApJ...172..165U} and \citet{1980A&A....91..175K} (which effectively only differ by a multiplicative constant $C_M$), they were able to reproduce the observations provided that constant factor is taken to be $C_M \sim 1000$ as in the \citet{1972ApJ...172..165U} model. Unfortunately, this is now known to be inconsistent with the fingering mixing rates measured in Direct Numerical Simulations (DNSs), which all agree that $C_M \sim 10$ at most \citep{2010ApJ...723..563D, 2011ApJ...728L..29T,2013ApJ...768...34B}. In other words, the basic state of fingering convection mixes chemical species at a rate that is two orders of magnitude smaller than what is required to explain RGB star observations.

More recent work by \citet{2015ApJ...808...89G} investigated the possibility that large-scale gravity waves or thermocompositional staircases might spontaneously form in fingering regions \citep[see][]{2013ApJ...768...34B}, which could cause an enhancement in transport as they do in the tropical ocean \citep{Schmitt685}. However, they concluded that this could not happen in the parameter regime appropriate for RGB stars. \citet{2018ApJ...862..136S} later proposed that including the effect of rotation might be the solution to the problem. Rotating fingering convection in the regime appropriate for RGB stars can give rise to the spontaneous formation of large-scale vortices, that greatly enhance transport by channeling vertical flows. However, that possibility bears a number of caveats, including the fact that vortex formation in rotating DNSs appears to depend on the horizontal aspect ratio of the domain \citep{julien_knobloch_plumley_2018} and the latitude of the region considered \citep{2017ApJ...834...44M}. Whether rotation is indeed the solution to the RGB problem therefore remains to be confirmed. 

In this Letter, we investigate an alternative possibility, and account for the effects of magnetic fields. We focus our attention on the effect of vertical fields (i.e., fields aligned with the direction of gravity). Section  \ref{sec:model} presents the underlying mathematical model, Section \ref{sec:stability} presents a brief linear stability analysis, and Section  \ref{sec:num} presents our numerical results. As we shall demonstrate, vertical magnetic fields rigidify the fingers and greatly enhance their vertical transport properties. These results also hold (at least qualitatively) in the inclined field case. Section \ref{sec:disc} discusses their implications for RGB star observations, and more generally in stellar astrophysics.

\section{Mathematical Model} \label{sec:model}
We consider a small stellar region whose vertical extent is smaller than a pressure scale height, which allows us to use the Boussinesq approximation for gases \citep{1960ApJ...131..442S}. Using a Cartesian domain, whose $z$-axis lies in the vertical ($z$) direction, the governing equations are: 
 \begin{align}
   &\nabla \cdot \bm{u} = 0 \label{eq:cont} \\[5pt]
  \begin{split}\label{eq:momentum}
   \rho_m \Big(\frac{\partial \bm{u}}{\partial t} + \bm{u} \cdot \nabla \bm{u} \Big) &= - \nabla p  +  \rho_m \nu \nabla ^2 \bm{u} \\[5pt] 
          &+ \frac{1}{\mu_0}(\nabla \times \bm{B}) \times \bm{B} +  \rho\bm{g} 
  \end{split}\\[5pt]
  &\frac{\partial T}{\partial t} + \bm{u} \cdot \nabla T - u_z \frac{dT_{\rm ad}}{dz} = \kappa_T \nabla^2 T \label{eq:temp} \\[5pt]
  &\frac{\partial C}{\partial t} + \bm{u} \cdot \nabla C = \kappa_C \nabla^2 C \label{eq:comp} \\[5pt]
  &\frac{\partial \bm{B}}{\partial t}  = \nabla \times (\bm{u} \times \bm{B}) + \eta \nabla ^2 \bm{B} \label{eq:induction}\\[5pt]
  &\nabla \cdot \bm{B} = 0 \label{eq:divergence}
 \end{align}
where $\bm{u} = (u_x, u_y, u_z)$ is the fluid velocity, $\bm{B} = (B_x, B_y, B_z)$ is the magnetic field, $\rho_m$ is the mean density of the fluid, $\bm{g}$ is the gravitational acceleration, $p$ is the pressure, $T$ is the temperature field, and $C$ is the compositional field, which can either represent the mean molecular weight of the fluid, or the concentration of a particular chemical species. The vertical adiabatic temperature gradient $\frac{dT_{\rm ad}}{dz}$ is equal to $-\frac{g}{c_p}$, where $c_p$ is the specific heat at constant pressure of the fluid. The kinematic viscosity $\nu$, and the thermal, compositional, and magnetic diffusivities, $\kappa_T$, $\kappa_C$, and $\eta$, respectively, are assumed to be constant. Here we assume that the magnetic permeability $\mu_0$ is simply that of a vacuum, which is equal to $4\pi$ in cgs units.

The equation of state is assumed to be linear over the domain, such that density perturbations $\rho'$ with respect to the mean density $\rho_m$ satisfy 
\begin{equation}
\frac{\rho'}{\rho_m} = -\alpha T' + \beta C',
\end{equation}
where $T'$ and $C'$ are perturbations to their respective background fields, such that $T = T_m + T'$ and $C = C_m + C'$, where $T_m$ and $C_m$ are constant. This model ignores the effects of magnetic buoyancy. The coefficients $\alpha$ and $\beta$ are constants of thermal expansion and compositional contraction, respectively, given by
\begin{equation}
\alpha = - \frac{1}{\rho_m} \frac{\partial \rho}{\partial T} \bigg\vert_{p, C}, \; \; \; \; \;
\beta = \frac{1}{\rho_m} \frac{\partial \rho}{\partial C} \bigg\vert_{p, T}.
\end{equation}

For both the thermal and compositional fields, we then assume the existence of a constant background gradient along the vertical direction, such that
\begin{align}
T' = z \frac{dT_0}{dz} + \tilde T, \\
C' = z \frac{dC_0}{dz} + \tilde C.
\end{align}

In all that follows, we shall assume that $\tilde T$, $\tilde C$, $\bm u$, and $\bm B$ are triply-periodic in the domain. The standard non-dimensionalization chosen for fingering convection is based on the length scale $d$ associated with the width of the fingers, given by \citep{doi:10.1111/j.2153-3490.1960.tb01295.x}
\begin{equation}
d = \Bigg( \frac{\kappa_T \nu}{\alpha g |\frac{d T_0}{dz} - \frac{dT_{\rm ad}}{dz}|} \Bigg) ^{1/4} = \Bigg( \frac{\kappa_T \nu}{N_T^2} \Bigg) ^{1/4},
\end{equation}
where $N_T$ is the local buoyancy frequency based on the temperature stratification only. The non-dimensional units for time and the remaining physical variables are then
\begin{flalign}
[t] &= \frac{d^2}{\kappa_T},  \hspace{15pt} [u] = \frac{\kappa_T}{d}, \hspace{15pt} [T] = d \, \bigg( \frac{d T_0}{dz} - \frac{dT_{\rm ad}}{dz} \bigg), \label{eq:nd1}\\[5pt]
 [C] &= \frac{\alpha}{\beta}d \, \bigg( \frac{d T_0}{dz} - \frac{dT_{\rm ad}}{dz} \bigg),  \hspace{13pt} [B] = B_0, \label{eq:nd2}
\end{flalign}
where $B_0$ is the reference magnetic field strength. Carrying these assumptions into the equations, our final, non-dimensionalized system describing fingering convection in the presence of magnetic fields is thus given by
\begin{align}
\begin{split}\label{eq:momentum_nd}
 \frac{\partial \hat{\bm{u}}}{\partial t} + \hat{\bm{u}} \cdot \nabla \hat{\bm{u}} &= - \nabla \hat{p} + \mathrm{Pr} \, \nabla ^2 \hat{\bm{u}} \\
   &+  H_B(\nabla \times \hat{\bm{B}}) \times \hat{\bm{B}} + \mathrm{Pr} \, ( \hat{T} - \hat{C})\hat{\bm{e}}_z,
 \end{split}\\[5pt]
& \frac{\partial \hat{T}}{\partial t} + \hat{\bm{u}} \cdot \nabla \hat{T} + \hat{u}_z  = \nabla^2 \hat{T}, \label{eq:temp_nd} \\[5pt]
& \frac{\partial \hat{C}}{\partial t} + \hat{\bm{u}} \cdot \nabla \hat{C} + \frac{\hat{u}_z}{R_0} = \tau \nabla^2 \hat{C}, \label{eq:comp_nd} \\[5pt]
& \frac{\partial \hat{\bm{B}}}{\partial t}  = \nabla \times (\hat{\bm{u}} \times \hat{\bm{B}}) + D_B \nabla ^2 \hat{\bm{B}}, \label{eq:induction_nd} \\[5pt]
& \nabla \cdot \hat{\bm{u}} = 0, \; \; \; \; \; \nabla \cdot \hat{\bm{B}} = 0, \label{eq:divergence_nd}
\end{align}
where from here onwards, hatted quantities denote non-dimensional ones and time and space variables have implicitly been made non-dimensional as well. In Eq.\ \eqref{eq:momentum_nd}, $\hat{\bm{e}}_z$ denotes the unit vector in the $z$ direction. 

The non-dimensional parameters controlling the system are the Prandtl number Pr, the compositional and magnetic diffusivity ratios $\tau$ and $D_B$, respectively, the density ratio $R_0$, and the Lorentz force coefficient $H_B$:
\begin{align}
\begin{split} \label{eq:nd_params}
&\mathrm{Pr} = \frac{\nu}{\kappa_T}, \; \tau = \frac{\kappa_C}{\kappa_T}, \; D_B = \frac{\eta}{\kappa_T} \\
&R_0 = \frac{\alpha \Big|\frac{d T_0}{dz} - \frac{dT_{\rm ad}}{dz}\Big|}{\beta \frac{d C_0}{dz}}, \\
&H_B = \frac{B_0^2 d^2}{\rho_m \mu_0 \kappa_T^2}.
\end{split}
\end{align}
 In the hydrodynamic limit ($H_B = 0$) the density ratio characterizes the stability of the system. Indeed, as shown by \citet{doi:10.1111/j.2153-3490.1960.tb01295.x}, a fluid is fingering unstable provided $1 < R_0 < 1/\tau$. \edit1{The parameter $H_B$ is the square of the ratio of the Alfv\'{e}n velocity to the characteristic finger velocity $\kappa_T/d$ used to non-dimensionalize the equations.} In both the analytical work and the numerical simulations that follow, we will vary this parameter with the goal of testing a variety of field strengths. We now briefly discuss what ranges of $H_B$ values we might expect in stellar fingering convection. 
 
 The typical values of $B_0$ that are likely to occur in stellar interiors can vary widely both within a given star and between different stars, depending on the stellar region under consideration. The same is true for the local conditions of the fluid (and thus the values of the other physical parameters). For example, fingering convection occurs in RGB stars at the base of the convective zone, while in main sequence (MS) and white dwarf (WD) stars it would occur near the surface following accretion of planets or debris. We provide order-of-magnitude estimates of $\nu$, $\kappa_T$, $\rho_m$, and $d$ in Table \ref{tab:params}, for the regions of MS stars, RGB stars, and WD stars where fingering convection could occur. \edit1{We see for instance that $d$ is indeed always much smaller than the pressure scale height $H_p$ (hence the justification of the Boussinesq approximation).} Based on these parameters, we can compute order-of-magnitude prefactors for $H_B$ for these three scenarios, getting
\begin{align}
    H_B^{\mathrm{MS}}  &\simeq 10^{-2} \Big( \frac{B_0}{100} \Big) ^2\Big( \frac{0.1}{\rho_m} \Big)^2 \Big( \frac{d}{10^4} \Big)^2 \Big( \frac{10^7}{\kappa_T} \Big)^2 \\ 
    H_B^{\mathrm{RGB}}  &\simeq 10^{-7}  \Big( \frac{B_0}{100} \Big) ^2\Big( \frac{1}{\rho_m} \Big)^2 \Big( \frac{d}{10^4} \Big)^2 \Big( \frac{10^9}{\kappa_T} \Big)^2 \label{eq:CL1} \\ 
    H_B^{\mathrm{WD}}  &\simeq 10^{-3} \Big( \frac{B_0}{100} \Big) ^2\Big( \frac{1}{\rho_m} \Big)^2 \Big( \frac{d}{10} \Big)^2 \Big( \frac{10^4}{\kappa_T} \Big)^2 \label{eq:CL2}
\end{align}
where all terms in the brackets are provided in cgs units.

We can therefore expect $H_B \ll 1$ as long as field strengths are limited to a few hundred G. While this may naively appear to imply that magnetic effects are irrelevant in stars, we will demonstrate in Section \ref{sec:disc} below that even for situations where $H_B$ is small, the effects of vertical background magnetic fields can be significant nonetheless. \edit1{This is because, at stellar parameters, the actual finger velocities are much smaller than $\kappa_T/d$ \citep{doi:10.1146/annurev-fluid-122316-045234}, so even weak fields can affect them.}

 \begin{table*}
\begin{center}
\begin{tabular}{ l l l l l l l}
\hline \hline \noalign{\smallskip}
Star type & $\kappa_T$ & $\nu$  & $\rho_m$ & $d=\Big(\frac{\kappa_T \nu}{N_T^2}\Big)^{1/4}$ & \edit1{$H_p$}\\
\hline \hline
\noalign{\smallskip}
Main Sequence &  $10^{7}$ & $10$  & $0.1$ & $10^{3.5} - 10^4$ & \edit1{$10^{10}$}\\
Red Giant Branch & $10^{9}$ & $100$   & $100-0.1$ & $10^{3.5} - 10^{4.5}$ & \edit1{$10^{10}$}\\
White Dwarf & $10^{2} - 10^{6}$ & $10-100$  & $10 - 0.1$ & $10^{0.5} - 10^{1.5}$ & \edit1{$10^{5}$}\\
\hline
\end{tabular}
\end{center}
\caption{Order-of-magnitude estimates for various governing parameters within the expected fingering convection regions of Main Sequence stars, Red Giant Branch stars, and White Dwarfs. The ranges represent values from the lower radius to the upper radius of the fingering region. \edit1{Note how, in all cases, $d \ll H_p$.} All units are in cgs.}
\label{tab:params}
\end{table*}

 In all that follows, we assume the existence of a uniform background field aligned with the vertical direction, and consider perturbations $\hat{\bm{ b}}$ to this background, so
  \begin{equation}
     \hat{\bm{B}} = \hat{\bm{e}}_z + \hat{\bm{b}},
 \end{equation}
\edit1{This model is consistent since we anticipate that any ambient magnetic field in a stellar radiative zone would vary on length scales that are much larger than the finger scale $d$, which is at most $\sim$100 m (see Table 1).} The case of arbitrarily inclined background fields will be studied in detail elsewhere (Harrington \& Garaud, in prep). We now proceed first to analyze the stability of the fluid to infinitesimal perturbations using linear stability analysis, then study the nonlinear saturation of the fingering instability in the presence of a large-scale vertical field in Section \ref{sec:num}.
 
 \section{Linear Stability} \label{sec:stability}
 We study the linear stability of the system in Eqs.\ \eqref{eq:momentum_nd} - \eqref{eq:divergence_nd}, using triply-periodic boundary conditions. Then, as in the analysis for hydrodynamic homogeneous fingering convection \citep[e.g.,][]{baines_gill_1969}, perturbations must be of the form
 \begin{equation}
     \hat{q} \propto \exp (\hat{\lambda} t + i(\hat{\bm{k}} \cdot \bm{x}))
 \end{equation}
for  $\hat{q} \in \{\hat{\bm{u}}, \, \hat{\bm{b}}, \, \hat{T}, \, \hat{C}\}$, where $\hat{\lambda}$ is the growth rate of each mode, $\hat{\bm{k}}= (\hat{k}_x, \hat{k}_y, \hat{k}_z)$ is the wave vector, and $\bm{x}=(x,y,z)$. 

Linearizing the governing equations assuming the perturbations $\hat{q}$ are small, and substituting the ansatz above yields, after some algebra, an equation for the non-dimensional growth rate $\hat{\lambda}$ of the form
\begin{align} 
\begin{split}\label{eq:growth}
\hat{\lambda} + \mathrm{Pr} \hat{k}^2 &+ \frac{\hat{k}_z^2 H_B}{\hat{\lambda} + D_B \hat{k}^2} \\& = \frac{\mathrm{Pr} \, \hat{k}_h^2}{\hat{k}^2}  \Bigg( \frac{1}{R_0(\hat{\lambda} + \tau \hat{k}^2)} 
- \frac{1}{(\hat{\lambda} + \hat{k}^2)}\Bigg) , 
\end{split}
\end{align}
where $\hat{k}^2 = \hat{k}_x^2 + \hat{k}_y^2 + \hat{k}_z^2$, and $\hat{k}_h^2$ is the square of the horizontal wave number, defined by $\hat{k}_h^2 = \hat{k}_x^2 + \hat{k}_y^2$.

We can see immediately that in the hydrodynamic limit ($H_B=0$), the relation for ordinary homogeneous fingering convection \citep[e.g.,][]{baines_gill_1969} is recovered. Interestingly, the same is true for the elevator modes ($\hat{k}_z=0$), showing that the fastest growing fingering modes are unaffected by the presence of a vertical field. The region of parameter space unstable to fingering convection therefore remains $1 < R_0 < \tau^{-1}$ \citep[as in the non-magnetic case, see][]{doi:10.1111/j.2153-3490.1960.tb01295.x}. The only effect of increasing $H_B$, or equivalently, increasing the background magnetic field strength, is that modes with higher $|\hat{k}_z|$ are suppressed \citep[see, e.g.,][]{2007A&A...476L..29C}, but the fastest-growing modes, which have $\hat{k}_z=0$, remain unchanged.

\section{Numerical Simulations} \label{sec:num}

\subsection{Simulation Parameters}
To study the nonlinear saturation of the fingering instability, we have modified the triply-periodic, pseudo-spectral PADDI code \citep{stellmach_traxler_garaud_brummell_radko_2011, 2011ApJ...728L..29T} to include magnetic fields and solve Eqs.\ \eqref{eq:momentum_nd} - \eqref{eq:divergence_nd}. The initial conditions for each simulation have the fluid completely at rest, under a uniform vertical magnetic field of unit strength (i.e., $\hat {\bm B} = \hat{ \bm{e}}_z$), and randomly generated small-amplitude perturbations in the $\hat{T}$ and $\hat{C}$ fields.

The simulations presented here have fixed values of the governing parameters in Eq. \eqref{eq:nd_params}, except for $H_B$. Realistic values for Pr, $\tau$, and $D_B$ are very small in stellar interiors, and are numerically unachievable with current technology, so we choose $\mathrm{Pr}=\tau=D_B=0.1$ in order to easily compare them with the existing body of non-magnetic simulations of fingering convection \citep[e.g.,][]{2011ApJ...728L..29T, 2013ApJ...768...34B,2018ApJ...862..136S}. The density ratio is chosen to be $R_0=1.45$, which lies in the region of parameter space fairly close to standard overturning convection. The suite of simulations presented here tests a wide range of background field strengths, with $H_B \in \{0.01, \, 0.1, \, 1, \, 10, \, 100\}$.

Each simulation has a spectral resolution of $96^3$ Fourier modes in each coordinate direction (which corresponds to an effective spatial resolution of $288^3$ mesh points) for a cubic domain of size $(100d)^3$, except for the $H_B=100$ run, which quadrupled the vertical extent of the simulation box (and proportionally reduced the length of the $y$-direction) in order to accurately model the highly elongated fingers which arise in that case.

\subsection{Qualitative Results}
\begin{figure*}
\plotone{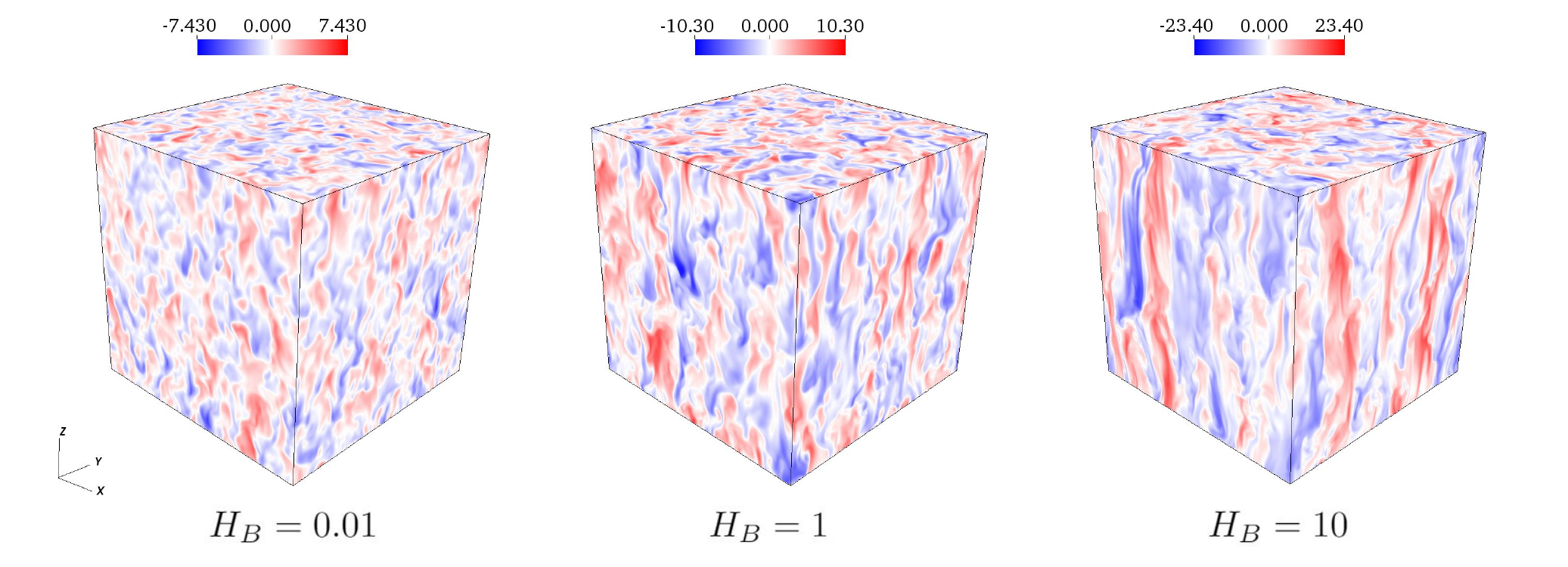}
\caption{Visualizations of the vertical fluid velocity component during the post-saturation, statistically stationary state, for runs with  $H_B=0.01$ (left), $H_B=1$ (middle), and $H_B=10$ (right). Increasing the strength of the vertical magnetic field (via $H_B$) imparts greater vertical coherence to the fingering structures.}
\label{fig:DNS_cubevis}
\end{figure*}

\begin{figure}
\plotone{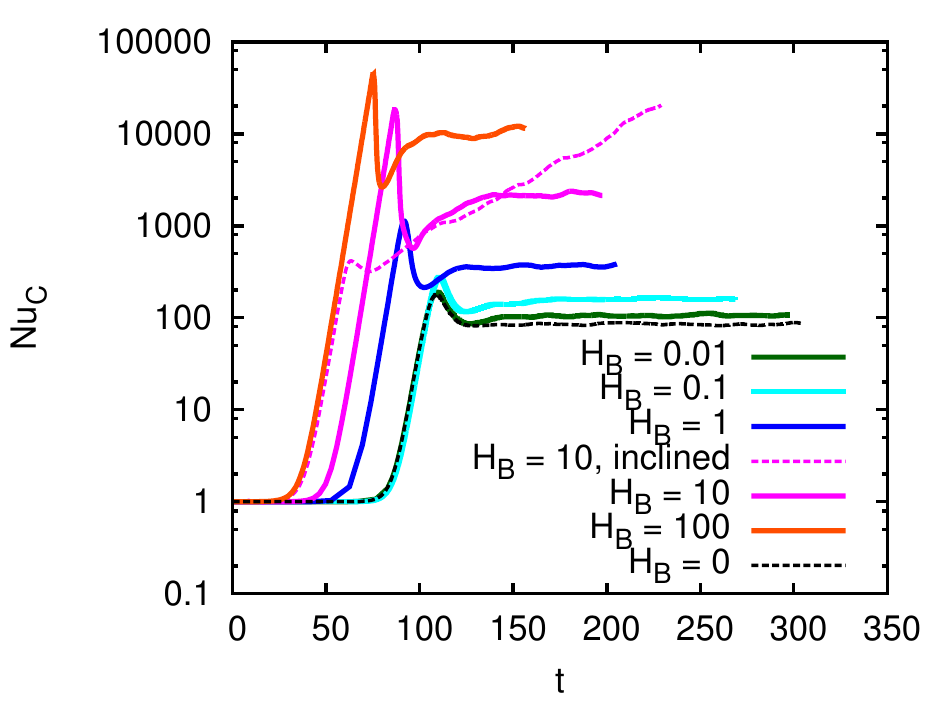}
\caption{The compositional Nusselt number $\mathrm{Nu}_C$ (see Eq.\ \eqref{eq:Nusselt}) as a function of time (in units of the thermal diffusion timescale) in simulations with varying $H_B$. The dashed purple line shows a case with a background field inclined at 45$^{\circ}$ from the $z$-axis, with $H_B=10$, and the black dashed-line shows a non-magnetic simulation ($H_B=0$).}
\label{fig:compNusselt}
\end{figure}

The qualitative properties of magnetized fingering convection are illustrated in Figures \ref{fig:DNS_cubevis} and \ref{fig:compNusselt}. As expected from linear theory, the instability initially grows exponentially at a rate that is independent of $H_B$, and eventually settles into a quasi-steady, weakly-turbulent state of small-scale fingering convection. As the strength of the background magnetic field increases via $H_B$, the fingers become more coherent and elongated along the vertical direction (see Figure \ref{fig:DNS_cubevis}). At the same time, the temperature and compositional fluxes, as well as the r.m.s.\ vertical velocity, all increase significantly (see Figure \ref{fig:compNusselt}).

Figure \ref{fig:DNS_cubevis} shows visualizations of the vertical component of the fluid velocity once the fingering convection is in a statistically steady state. The $H_B = 0.01$ case is indistinguishable from the non-magnetic case, with fingers that have a roughly unit aspect ratio. As $H_B$ increases, we see an increasing anisotropy of the fingers, which become coherent over long vertical distances, as well as a marked increase in their vertical velocities.

Qualitatively speaking, this can be explained by noting that increasing the field strength rigidifies the initial fingers vertically and delays saturation until a much higher r.m.s.\ vertical fluid velocity is reached. This increase in the vertical velocities within the fingers causes a substantial increase in the vertical turbulent compositional fluxes, as measured by the compositional Nusselt number $\mathrm{Nu}_C$, defined by
\begin{equation}
    \mathrm{Nu}_C = 1 - \frac{R_0}{\tau} \langle \hat{u}_z \hat{C} \rangle = \frac{D_C}{\kappa_C}, \label{eq:Nusselt}
\end{equation}
where $\langle \, \rangle$ denotes a volume average over the domain and $D_C$ is the effective compositional diffusivity (i.e., the sum of the microscopic plus turbulent one). Figure \ref{fig:compNusselt} shows the evolution of $\mathrm{Nu}_C$ over time for each of the simulations. We see that stronger field strengths can significantly enhance the compositional transport by up to a few orders of magnitude compared with the non-magnetic case.

Also plotted in Fig.\ \ref{fig:compNusselt} is the evolution of $\mathrm{Nu}_C$ for a $H_B=10$ simulation, with a background magnetic field inclined at 45$^{\circ}$ from the $z$-axis. The behavior for arbitrarily inclined background fields is more complex and is thus saved for a later work (Harrington \& Garaud, in prep.), but preliminary results such as this one indicate that equally significant enhancements of compositional transport rates are not just attainable, but are to be expected.

\section{Quantitative Analysis} \label{sec:disc}

We now provide a simple quantitative model for the increase in thermocompositional fluxes caused by the presence of a vertical field. Previous work has shown that the mechanism responsible for saturation of ordinary fingering convection is the development of a shear instability between adjacent up-flowing and down-flowing fingers \citep{2012JFM...692....5R, 2013ApJ...768...34B}, so an obvious explanation for our results is that the vertical magnetic field suppresses the shear instability. We now revisit the \citet{2013ApJ...768...34B} model, and include the effects of a vertical field.

In the hydrodynamic limit, \citet{2013ApJ...768...34B} assume that the fingers saturate when the growth rate $\hat{\sigma}$ of shear instabilities between up- and down-flowing fingers becomes commensurate with the growth rate $\hat{\lambda}_f$ of the fastest-growing modes of the basic fingering instability. That problem can be solved analytically using dimensional analysis, since the growth rate of the shearing instability must be $\hat{\sigma} \propto \hat{w}_f \hat{l}_f$ where $\hat{w}_f$ is the velocity in the fingers, and $\hat{l}_f$ is their horizontal wavenumber. Assuming that $\hat{\lambda}_f = C_B \hat{\sigma} = C_B \hat{w}_f \hat{l}_f$, where $C_B$ is a universal constant, then provides an estimate for $\hat{w}_f$, namely $\hat{w}_f = \hat{\lambda}_f / C_B \hat{l}_f$. This was verified to hold by \citet{2018ApJ...862..136S}, who found that $C_B \approx \frac{1}{2\pi}$.

To compute $\mathrm{Nu}_C$, \citet{2013ApJ...768...34B} then assumed that 
\begin{equation}
    \langle \hat{u}_z \hat{C} \rangle \approx - K_B \, \frac{\hat{w}_f^2}{R_0 (\hat{\lambda} + \tau \hat{l}_f^2)},
\end{equation}
where $K_B$ is another constant and is of order unity. This then yields the formula
\begin{equation}
    \mathrm{Nu}_C = 1 + K_B \frac{ \hat{w}_f^2}{\tau(\hat{\lambda} + \tau \hat{l}_f^2)} = 1 + \frac{K_B}{C_B^2} \frac{ \hat{\lambda}_f^2}{\tau \hat{l}_f^2(\hat{\lambda} + \tau \hat{l}_f^2)}, \label{eq:Nu2}
\end{equation}
which was fitted against data from numerical simulations to find that $\frac{K_B}{C_B^2} \approx 49$, which means $K_B \simeq 1.24$.

\begin{figure}
\plotone{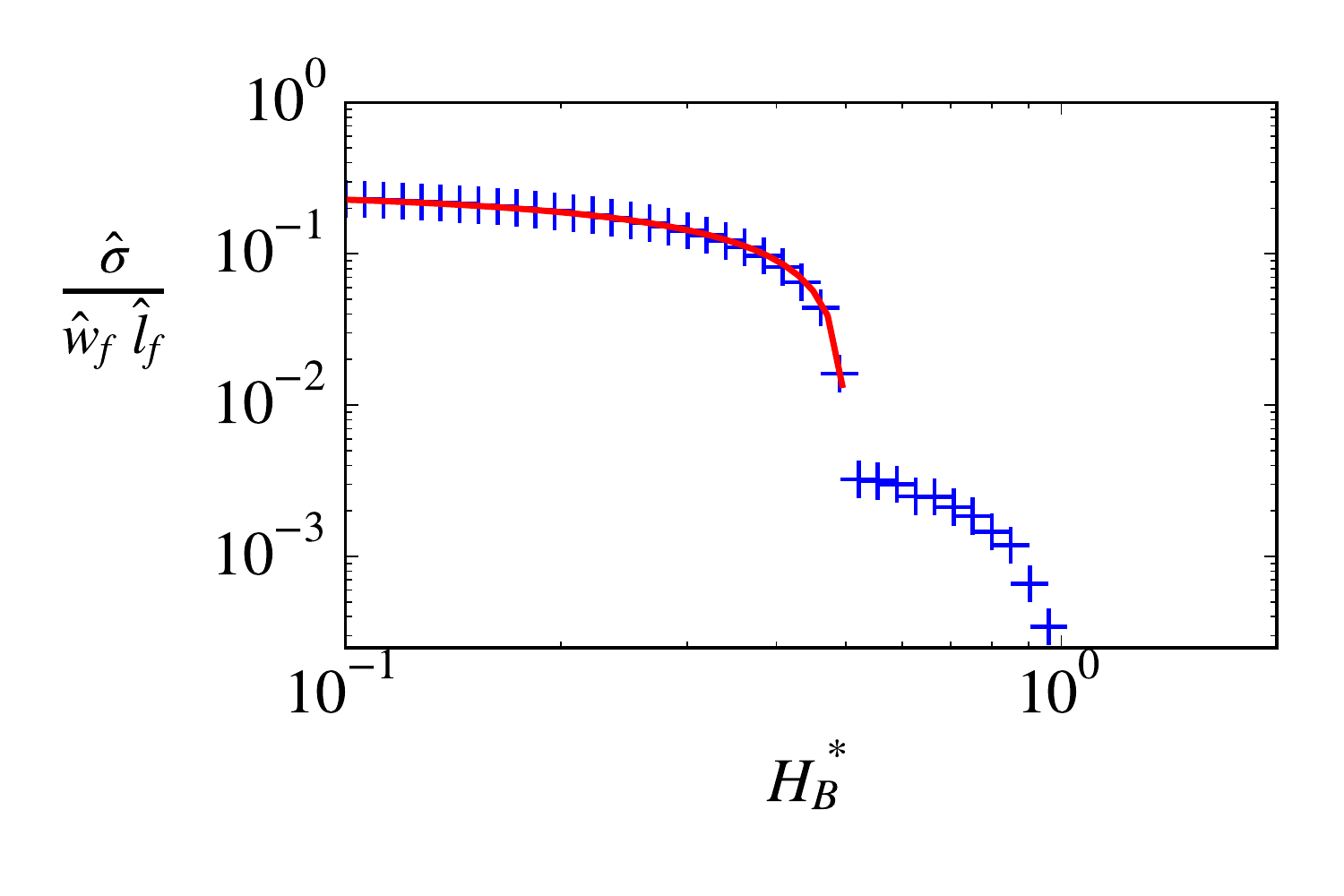}
\caption{The non-dimensional growth rate of the shear instability $\hat{\sigma}$ as a function of $H_B^*$ (blue crosses). The red line shows the fit given by Eq.\ \eqref{eq:shearfit}.}
\label{fig:shear}
\end{figure}

A vertical magnetic field, on the other hand, stabilizes the fingers against shear instabilities, so that larger velocities are required to trigger them. To see this, we studied formally the stability of a sinusoidal shear flow of the kind $\hat{w}_f \sin(\hat{l}_f x) \hat{\bm{e}}_z$ (which mimics the flow within the finger elevator modes) in the presence of a constant vertical field of unit amplitude, by extending the Floquet analysis of \citet{2013ApJ...768...34B} (see their Appendix A). While the details of this calculation are presented elsewhere (Harrington, MS Thesis), the results are shown in Figure \ref{fig:shear}. We find that the growth rate of the shear instability $\hat{\sigma}$ now depends sensitively on the non-dimensional number
\begin{equation} \label{eq:H_B*}
   H_B^* =\frac{H_B}{\hat{w}_f^2},
\end{equation}
which decreases as the velocity in the fingers increases.

There are two sets of modes unstable to shear -- a slowly growing one, destabilized for $H_B^* < 1$, and a rapidly growing one, destabilized for $H_B^* <0.5$. \edit1{The relevant threshold is $H_B^* = 0.5$, which corresponds to equipartition between the kinetic energy of the fingers and the magnetic energy of the background field (i.e., dimensionally speaking, $H_B^* = 0.5$ implies $\frac{1}{2} \rho_m \langle u_z \rangle ^2 = \frac{B_0^2}{2\mu_0}$). For $H_B^*$ below this threshold, the shear modes have a vertical wave number of $0.2 \hat l_f$ - $0.6 \hat l_f$, which implies that the vertical wavelength of the instability is about 2-5 times as long as the horizontal wavelength of the fingers.}  We fit the branch with larger growth rate as a function of $H_B^*$, getting
\begin{equation}\label{eq:shearfit}
\frac{\hat{\sigma}}{\hat{w}_f \hat{l}_f} \simeq 0.42 (0.5 - H_B^*)^{2/3}.
\end{equation}
As in the \citet{2013ApJ...768...34B} model, we then assume that \edit1{saturation occurs when} $\hat{\sigma}$ is of the order of the growth rate of the fingers $\hat{\lambda}_f$, according to
\begin{equation}\label{eq:shearrelation}
    0.42 \hat{w}_f \hat{l}_f (0.5 - H_B^*)^{2/3} = C_H \hat{\lambda}_f,
\end{equation}
where $C_H$ is a universal constant. By demanding that the $H_B^* = 0$ (hydrodynamic) limit reproduces the proportionality relation $C_B \hat{w}_f \hat{l}_f = \hat{\lambda}_f$, we determine that $C_H=(0.42)(0.5^{2/3})/C_B \approx 1.66$. Combining Eqs.\ \eqref{eq:H_B*} and \eqref{eq:shearrelation}, we can then express $\hat{w}_f$ in terms of $H_B$, yielding a relation that is quartic in $\hat{w}_f^{1/2}$:
\begin{equation}
    0.5 \, \hat{w}_f^2 - H_B = \Big(C_H \frac{\hat{\lambda}_f}{0.42 \, \hat{l}_f} \Big)^{3/2} \hat{w}_f^{1/2}. \label{eq:w_relation}
\end{equation}

\begin{figure*}
\plottwo{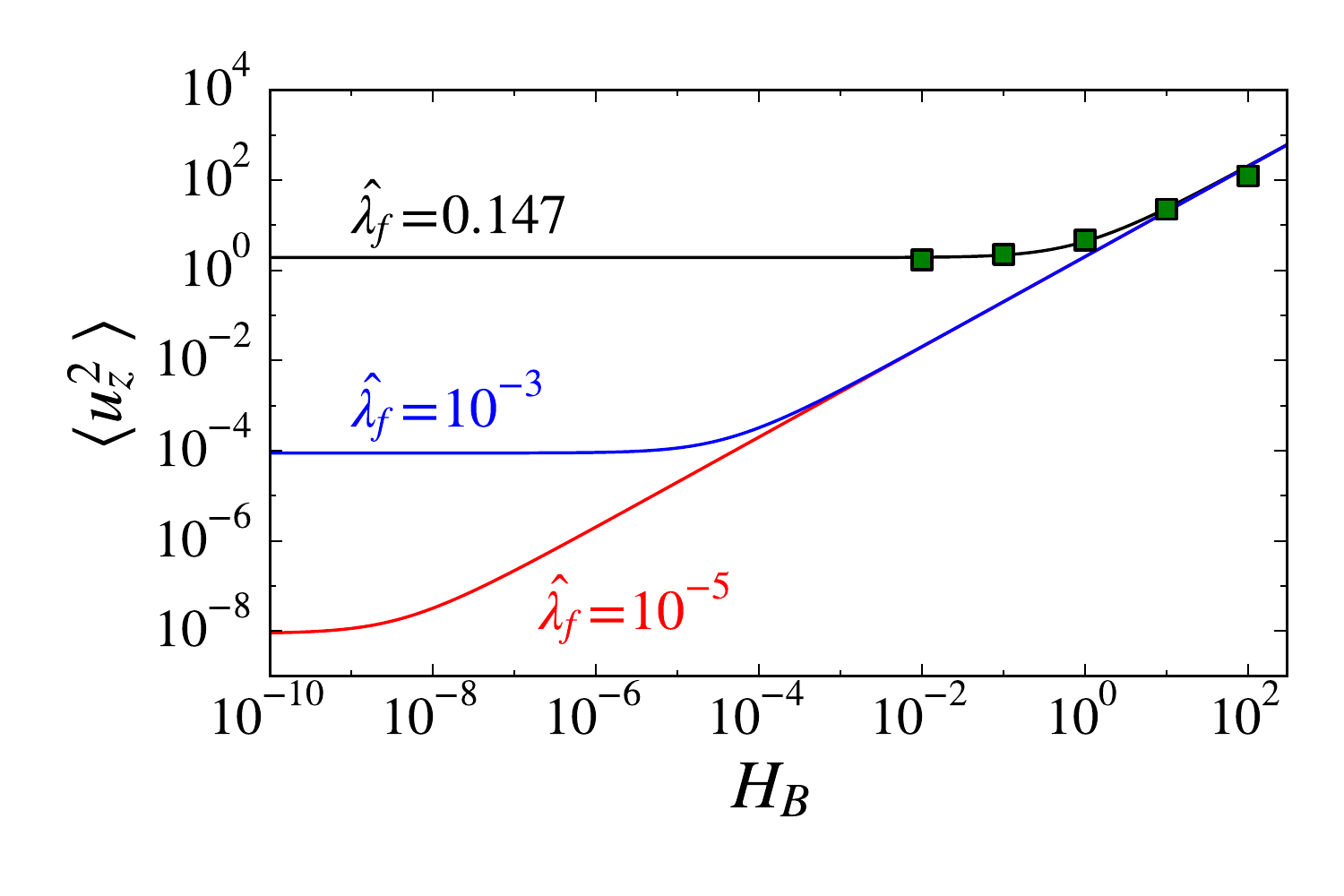}{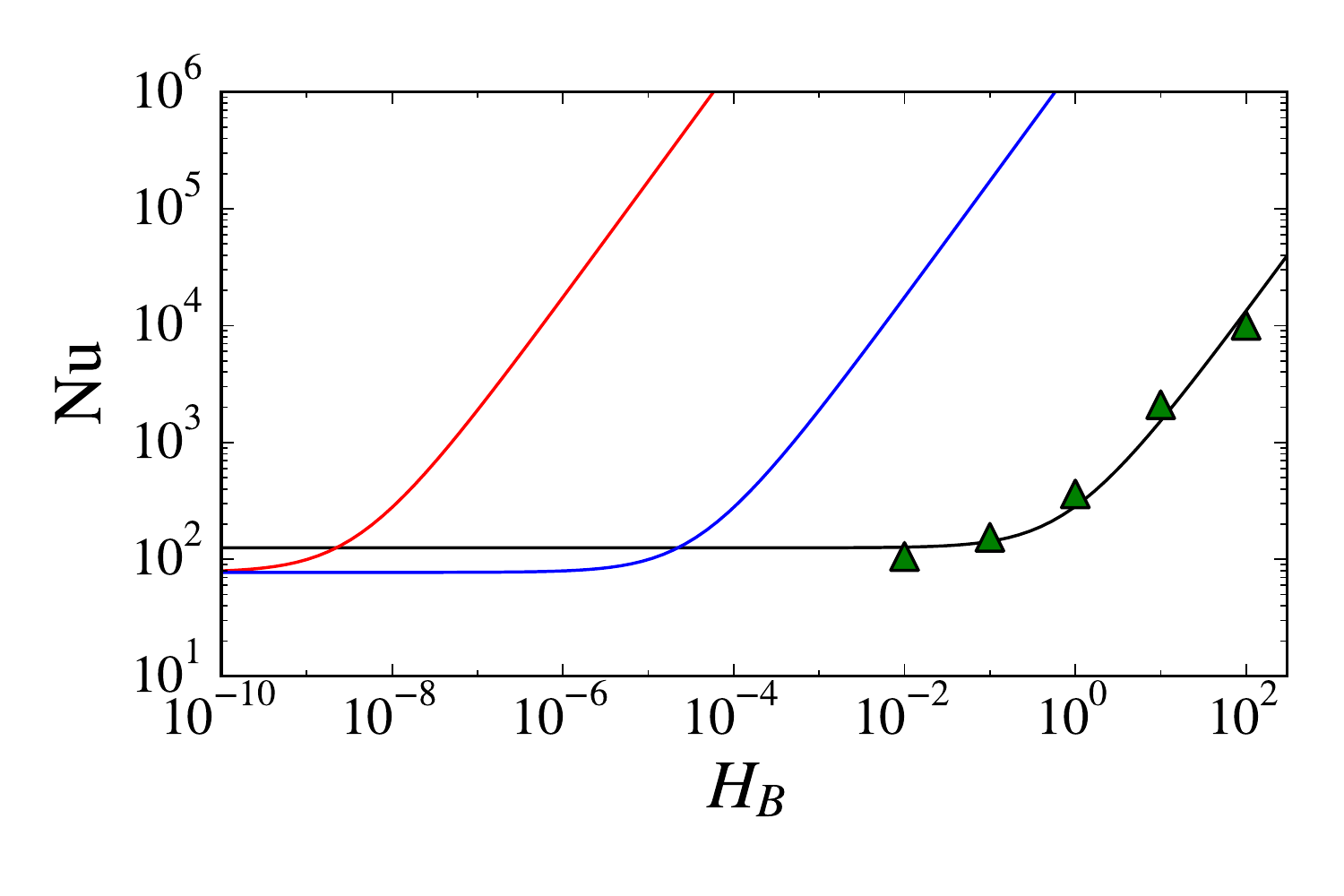}
\caption{The volume-averaged squared vertical velocity $\langle u_z^2 \rangle$ (left, green squares) and compositional Nusselt number Nu$_C$ (right, green triangles) as a function of $H_B$ from the simulations. The different lines show the prediction from Eq.\ \eqref{eq:w_relation} for $\hat{\lambda}_f=0.147$ (black), corresponding to the parameters in the simulations; $\hat{\lambda}_f=10^{-3}$ (blue), representative of a white dwarf's fingering convection environment; and $\hat{\lambda}_f=10^{-5}$ (red), representative of an RGB star's fingering convection environment.}
\label{fig:scalings}
\end{figure*}

We can immediately see two asymptotic regimes arising from this relation. The first is for very small $H_B$, where the velocity in the fingers simply approaches that of the \citet{2013ApJ...768...34B} hydrodynamic model. However, for very large $H_B$, the RHS term becomes negligible and the velocity in the fingers behaves roughly as $\hat{w}_f = \sqrt{2H_B}$\edit1{, which, as discussed earlier, corresponds to an exact equipartition between the magnetic energy of the background field and the kinetic energy of the fingers.} We call this the ``magnetically-dominated" regime. The values of $H_B$ where the transition between the two regimes occurs thus depends on the growth rate $\hat{\lambda}_f$ and horizontal wave number $\hat{l}_f$ of the elevator modes, which are in turn dependent on the governing parameters ($\mathrm{Pr}$, $\tau$, $R_0$).

We can solve Eq.\ \eqref{eq:w_relation} numerically for $\hat{w}_f$ as a function of $H_B$ for various parameter values, the results of which are shown in Figure \ref{fig:scalings}. With $\mathrm{Pr}=\tau = D_B=0.1$, and $R_0=1.45$, as in the numerical simulations, we have $\hat{\lambda}_f\approx0.147$ and $\hat{l}_f\approx0.666$, and find that the numerical results for $\langle \hat{u}_z ^2 \rangle$ are well predicted by $\hat{w}_f^2$ computed from Eq.\ \eqref{eq:w_relation}. We can see that the transition between the low- and high-$H_B$ regimes for these parameter values occurs around $H_B=1$.

However, in stellar interiors, the Prandtl number Pr (as well as $\tau$) can be several orders of magnitude smaller than what we are able to simulate numerically, and in this $(\mathrm{Pr}, \, \tau) \ll 1$ limit, we typically have \citep[see Appendix B of ][]{2013ApJ...768...34B}
\begin{equation}
    \hat{\lambda}_f \simeq \sqrt{ \frac{ \mathrm{Pr} \, \tau (1/\tau - 1)}{R_0 - 1}} \simeq  \sqrt{ \frac{ \mathrm{Pr}}{R_0 - 1}} \ll 1, \label{eq:lambda_scale}
\end{equation}
which means the transition between low- and high-$H_B$ regimes now occurs at a much smaller value of $H_B$. In Figure \ref{fig:scalings}, we have also solved Eq.\ \eqref{eq:w_relation} for $\hat{\lambda}_f=10^{-3}$ as well as $\hat{\lambda}_f=10^{-5}$ (keeping $\hat{l}_f = 0.666$ fixed since $\hat{l}_f$ remains $O(1)$ in the low-Pr limit), which are representative values of what we would expect in a WD or RGB star, respectively. These results show that the magnetically-dominated regime is $H_B \geq 10^{-4}$ for WD stars and $H_B \geq 10^{-8}$ for RGB stars. Thus, based on our estimates in Eqs.\ \eqref{eq:CL1} and \eqref{eq:CL2}, it is reasonable to expect that fingering convection in such stars can be significantly affected by magnetic fields.

Using our model for $\hat{w}_f^2$, we can finally compute the predicted turbulent compositional flux in magnetized fingering convection via 
\begin{equation}
    \mathrm{Nu}_C = 1 + K_B \frac{ \hat{w}_f^2}{\tau(\hat{\lambda} + \tau \hat{l}_f^2)},
\end{equation}
with the same value of $K_B$ as in \citet{2013ApJ...768...34B}. The results are summarized in the right panel of Figure \ref{fig:scalings}, which shows $\mathrm{Nu}_C$ as a function of $H_B$, for the same three parameter regimes (numerical simulations, WD stars, and RGB stars). 

We find that the value of $\mathrm{Nu}_C$ measured in the statistically stationary state in all of our simulations is well-predicted by our model. Crucially, we see that $\mathrm{Nu}_C$ scales like $H_B$ in the magnetically-dominated regime, which can easily be understood since $\mathrm{Nu}_C \propto \hat{w}_f^2 \propto H_B$ in that case. This means that $\mathrm{Nu}_C$ can increase by orders of magnitude depending on the background field strength. In fact, using Eqs.\ \eqref{eq:lambda_scale} and \eqref{eq:Nu2}, together with the definitions of $d$ and $H_B$, our model predicts that the turbulent compositional diffusivity due to magnetized fingering convection should be equal to
\begin{equation}
D_C \simeq 2 K_B \frac{B_0^2}{\rho_m \mu_0} \sqrt{ \frac{N_T^2 + N_\mu^2}{-N_T^2 N_\mu^2}}, \label{eq:DC}
\end{equation}
where $N^2_T = \alpha g (dT_0/dz -dT_{\mathrm{ad}}/dz) $ is the square of the temperature-based buoyancy frequency, $N_\mu^2 = - \beta g dC_0/dz$ is the square of the compositional buoyancy frequency (which is negative since the compositional field is destabilizing), and where we have assumed that $\hat \lambda_f \gg \tau \hat l_f$, which is typically the case for $R_0 \ll \mathrm{Pr}^{-1/2}$. Equation \eqref{eq:DC} should hold as long as we remain in the magnetically-dominated regime, which, as discussed earlier, corresponds to the limit $H_B \ge 10^{-4}$ in WDs, and $H_B \ge 10^{-8}$ in RGB stars. 

Finally, note that the enhancement in the vertical finger velocity by magnetic fields can also affect heat transport, which is normally negligible in hydrodynamic fingering convection \citep{2011ApJ...728L..29T}. We predict using similar arguments that the equivalent Nusselt number for (potential) temperature should be 
\begin{equation}
\mathrm{Nu}_T = 1 + K_B \frac{\hat{w}_f^2}{\hat{\lambda} + \hat{l}_f^2},
\end{equation}
with a corresponding dimensional heat flux given by
\begin{equation}
F_T = - \rho_m c_p \kappa_T \frac{dT_0}{dz} + \rho_m c_p \kappa_T \left( \frac{dT_0}{dz} - \frac{dT_{\rm ad}}{dz} \right) (1- {\rm Nu}_T).
\end{equation}
With $\mathrm{Nu}_T \gg 1$, we note the potential for transporting heat inward, as the right-hand term (which is usually small since $\mathrm{Nu}_T$ is ordinarily close to 1) can be made significantly negative.

\section{Discussion}
Our results have obvious implications for the RGB stars abundance problem \citep{2000A&A...354..169G}. Figure \ref{fig:scalings} shows that even a moderate magnetic field of $\sim 300$ G (for which $H_B \sim 10^{-6}$) would increase the value of the turbulent mixing coefficient by two orders of magnitude compared with the non-magnetic case, which would then be sufficient to explain the observations \citep[cf.][]{2007A&A...467L..15C}. Such magnetic field strengths are not unreasonably large, and would indeed be likely in RGB stars. Although our numerical results so far have been limited to the vertical field case, we have also shown that similar (or even larger) enhancements of the turbulent fluxes are likely if the field is inclined, so we expect our conclusions to be robust. 

Our results may also help solve another RGB-related ``missing mixing'' puzzle. Indeed, since the mixing coefficient $D_C$ depends on the magnetic field strength, which in turn most likely decreases with increasing radius within the star (e.g. if the field is of primordial origin, or was created by a dynamo in a prior core-convective phase), we predict that $D_C$ should decrease sharply with radius away from the hydrogen-burning shell. This might provide a more natural explanation for the radially varying mixing coefficient required to explain concurrent Li and C abundances in carbon-enhanced metal poor RGB stars \citep{2018ApJ...863L...5H}. 

Beyond RGB stars, we also predict that moderate magnetic fields could enhance fingering-induced mixing in WD (and MS) stars, an effect that should be taken into account when inferring the accretion rate of planetary debris onto the star, for instance.

\acknowledgments
P. H. and P. G. acknowledge funding by NSF-AST 1412951 and 1517927. Simulations were run on the UCSC Hyades cluster, purchased with an NSF MRI grant. The authors thank S. Sengupta, \edit1{S. Stellmach,} and J. Schwab for insightful discussions.

%\bibliography{refs}

\begin{thebibliography}{}
\expandafter\ifx\csname natexlab\endcsname\relax\def\natexlab#1{#1}\fi
\providecommand{\url}[1]{\href{#1}{#1}}

\bibitem[{Baines \& Gill(1969)}]{baines_gill_1969}
Baines, P.~G., \& Gill, A.~E. 1969, Journal of Fluid Mechanics, 37, 289–306

\bibitem[{{Brown} {et~al.}(2013){Brown}, {Garaud}, \&
  {Stellmach}}]{2013ApJ...768...34B}
{Brown}, J.~M., {Garaud}, P., \& {Stellmach}, S. 2013, \apj, 768, 34

\bibitem[{{Charbonnel} \& {Zahn}(2007{\natexlab{a}})}]{2007A&A...467L..15C}
{Charbonnel}, C., \& {Zahn}, J.-P. 2007{\natexlab{a}}, \aap, 467, L15

\bibitem[{{Charbonnel} \& {Zahn}(2007{\natexlab{b}})}]{2007A&A...476L..29C}
---. 2007{\natexlab{b}}, \aap, 476, L29

\bibitem[{{Denissenkov}(2010)}]{2010ApJ...723..563D}
{Denissenkov}, P.~A. 2010, \apj, 723, 563

\bibitem[{Garaud(2018)}]{doi:10.1146/annurev-fluid-122316-045234}
Garaud, P. 2018, Annual Review of Fluid Mechanics, 50, 275.
\newblock \url{https://doi.org/10.1146/annurev-fluid-122316-045234}

\bibitem[{{Garaud} {et~al.}(2015){Garaud}, {Medrano}, {Brown}, {Mankovich}, \&
  {Moore}}]{2015ApJ...808...89G}
{Garaud}, P., {Medrano}, M., {Brown}, J.~M., {Mankovich}, C., \& {Moore}, K.
  2015, \apj, 808, 89

\bibitem[{{Gratton} {et~al.}(2000){Gratton}, {Sneden}, {Carretta}, \&
  {Bragaglia}}]{2000A&A...354..169G}
{Gratton}, R.~G., {Sneden}, C., {Carretta}, E., \& {Bragaglia}, A. 2000, \aap,
  354, 169

\bibitem[{{Henkel} {et~al.}(2018){Henkel}, {Karakas}, {Casey}, {Church}, \&
  {Lattanzio}}]{2018ApJ...863L...5H}
{Henkel}, K., {Karakas}, A.~I., {Casey}, A.~R., {Church}, R.~P., \&
  {Lattanzio}, J.~C. 2018, ApJL, 863, L5

\bibitem[{Julien {et~al.}(2018)Julien, Knobloch, \&
  Plumley}]{julien_knobloch_plumley_2018}
Julien, K., Knobloch, E., \& Plumley, M. 2018, Journal of Fluid Mechanics, 837,
  R4

\bibitem[{{Kippenhahn} {et~al.}(1980){Kippenhahn}, {Ruschenplatt}, \&
  {Thomas}}]{1980A&A....91..175K}
{Kippenhahn}, R., {Ruschenplatt}, G., \& {Thomas}, H.-C. 1980, \aap, 91, 175

\bibitem[{{Moll} \& {Garaud}(2017)}]{2017ApJ...834...44M}
{Moll}, R., \& {Garaud}, P. 2017, \apj, 834, 44

\bibitem[{{Radko} \& {Smith}(2012)}]{2012JFM...692....5R}
{Radko}, T., \& {Smith}, D.~P. 2012, Journal of Fluid Mechanics, 692, 5

\bibitem[{Schmitt {et~al.}(2005)Schmitt, Ledwell, Montgomery, Polzin, \&
  Toole}]{Schmitt685}
Schmitt, R.~W., Ledwell, J.~R., Montgomery, E.~T., Polzin, K.~L., \& Toole,
  J.~M. 2005, Science, 308, 685.
\newblock \url{http://science.sciencemag.org/content/308/5722/685}

\bibitem[{{Sengupta} \& {Garaud}(2018)}]{2018ApJ...862..136S}
{Sengupta}, S., \& {Garaud}, P. 2018, \apj, 862, 136

\bibitem[{{Spiegel} \& {Veronis}(1960)}]{1960ApJ...131..442S}
{Spiegel}, E.~A., \& {Veronis}, G. 1960, \apj, 131, 442

\bibitem[{Stellmach {et~al.}(2011)Stellmach, Traxler, Garaud, Brummell, \&
  Radko}]{stellmach_traxler_garaud_brummell_radko_2011}
Stellmach, S., Traxler, A., Garaud, P., Brummell, N., \& Radko, T. 2011,
  Journal of Fluid Mechanics, 677, 554–571

\bibitem[{Stern(1960)}]{doi:10.1111/j.2153-3490.1960.tb01295.x}
Stern, M.~E. 1960, Tellus, 12, 172.
\newblock
  \url{https://onlinelibrary.wiley.com/doi/abs/10.1111/j.2153-3490.1960.tb01295.x}

\bibitem[{{Traxler} {et~al.}(2011){Traxler}, {Garaud}, \&
  {Stellmach}}]{2011ApJ...728L..29T}
{Traxler}, A., {Garaud}, P., \& {Stellmach}, S. 2011, \apjl, 728, L29

\bibitem[{{Ulrich}(1972)}]{1972ApJ...172..165U}
{Ulrich}, R.~K. 1972, \apj, 172, 165

\end{thebibliography}

%% This command is needed to show the entire author+affilation list when
%% the collaboration and author truncation commands are used.  It has to
%% go at the end of the manuscript.
%\allauthors

%% Include this line if you are using the \added, \replaced, \deleted
%% commands to see a summary list of all changes at the end of the article.
%\listofchanges

\end{document}